\documentclass[aps,prb,twocolumn]{revtex4}
\usepackage{epsfig,latexsym,amssymb,amsmath,amsbsy,graphics,graphicx}

\def \beq{\begin{equation}}
\def \eeq{\end{equation}}
\def \beqarr{\begin{eqnarray}}
\def \eeqarr{\end{eqnarray}}

\begin{document}

\title{Collective Modes and Skyrmion Excitations in Graphene SU(4) Quantum Hall Ferromagnets }

\author{Kun Yang}

\affiliation{National High Magnetic Field Laboratory and Department
of Physics, Florida State University, Tallahassee, FL 32306, USA}

\author{S. Das Sarma}

\affiliation{Condensed Matter Theory Center, Department of Physics,
University of Maryland, College Park, MA 20742, USA}

\author{A. H. MacDonald}

\affiliation{Department of Physics, University of Texas, Austin, TX
78712, USA}

\date{\today}

\begin{abstract}
Graphene exhibits quantum Hall ferromagnetism in which an
approximate $SU(4)$ symmetry involving spin and valley degrees of
freedom is spontaneously broken. We construct a set of integer and
fractional quantum Hall states that break the $SU(4)$ spin/valley
symmetry, and study their neutral and charged excitations. Several
properties of these ferromagnets can be evaluated analytically in
the $SU(4)$ symmetric limit, including the full collective mode
spectrum at integer fillings. By constructing explicit wave
functions we show that the lowest energy skyrmion states carry
charge $\pm 1$ for {\em any} integer filling, and that skyrmions are
the lowest energy charged excitations for graphene Landau level
index $|n| \le 3$. We also show that the skyrmion lattice states
which occur near integer filling factors support four gapless
collective mode branches in the presence of full $SU(4)$ symmetry.
Comparisons are made with the more familiar $SU(2)$ quantum Hall
ferromagnets studied previously.
\end{abstract}


\maketitle

\section{Introduction}

Recent experimental work\cite{graphene_exp} has established graphene
as a new two-dimensional (2D) electron system with linear Dirac-like
energy-band dispersion.  An important aspect of graphene physics is
the two-fold valley degeneracy of its low-energy band structure
which, in combination with the usual spin doublet, implies band
eigenstates with degeneracy $N = 4$.  In a strong magnetic field,
the four-fold degeneracy of Landau levels in graphene has been
clearly observed in recent quantum Hall
measurements.\cite{novoselov,zhang,zhang2}  This property is
expected\cite{nomura} to lead to an intriguing interplay between
interaction and quantum Hall physics by introducing a rich variety
of new states with different types of spontaneous symmetry breaking,
and new varieties of low-lying collective modes. In this paper we
explore some of the theoretical possibilities for quantum Hall
ground states and collective modes that follow from the enlarged
($N=4$) Landau level degeneracy of graphene, emphasizing
similarities and differences compared to the well-studied $N=2$,
spin-only circumstance\cite{sds,gm} relevant to GaAs 2D systems.

One of the most theoretically
interesting\cite{sondhi,fertig,skyrtheory} and phenomenologically
significant\cite{skyrexpt} consequences of quantum Hall
ferromagnetism is the presence of a finite density of skyrmions in
the ground state near integer filling factors. A dense system of
skyrmions constitutes an emergent set of low-energy degrees of
freedom that qualitatively alters\cite{skyrexpt} NMR, optical,
thermal and transport properties of the two-dimensional electron
system. In the $N=2$ case, dense skyrmion systems occur only near
Landau level filling factor $\nu=1$. We predict that in graphene
dense skyrmion states\cite{sondhi,fertig,skyrexpt} occur near many
integer filling factors, and that they have a larger number of
gapless collective modes than for $N=2$.\cite{skyrnmr} For Coulombic
electron-electron interactions, skyrmion lattice states occur
only\cite{sondhiwu} near $\nu=1$ in the $SU(2)$ case. For graphene
we predict skyrmion lattice states near $\nu = \pm \, l$ for all $l
< 14$ except for $l=2,6,10$ when all Landau-level multiplets are
either full or empty, and that skyrmion lattice states in graphene
have four branches of gapless collective modes at all these filling
factors in the absence of symmetry-breaking perturbations.

The $N=4$ internal degrees-of-freedom present in graphene take on
particular significance because of the very strong Coulomb
interaction energy scale (estimated to be more than 1000K at
45T,\cite{zhang2} several times larger than in GaAs) which will help
make all the physics that follows from interaction-driven,
spontaneous symmetry breaking ({\em i.e.} phases which break the
internal symmetry associated with the degeneracy $N$) more
accessible experimentally. The point of departure for our analysis
will be an $SU(N=4)$ symmetric Hamiltonian which allows us to obtain
a number of exact results.  This highly symmetric model is believed
to be a good approximation of the full Hamiltonian of
graphene.\cite{nomura,goerbig} For relatively low carrier densities
the largest symmetry breaking term in the Hamiltonian will be Zeeman
coupling which favors spin alignment along the field direction.  We
will discuss some of the important consequences of this term.

We remark that the possibility of a spontaneously broken valley
symmetry is akin to orbital ordering, with the novel twist that the
orbital ordering is in {\em momentum space}, rather than in real
space. The anticipated exotic orbital order in graphene should
co-exist with spin order making graphene (along with systems such as
superfluid He-3, manganites, and related systems where real space
orbital order apparently coexists with spin order) an interesting
system to study novel quantum phases with interplay between spin and
orbital order. The momentum space orbital ordering discussed here
may also imply observable Jahn-Teller coupling effects in graphene,
but we do not explore this idea further in this paper.

The consequences of $SU(4)$ or more generally, $SU(N)$ symmetry in
quantum Hall physics have been discussed previously by Arovas and
co-workers\cite{arovas} in the context of Silicon systems which also
have additional (approximate) valley degeneracies, and by Ezawa and
co-workers\cite{ezawa03,ezawa05} in the context of bilayer quantum
Hall systems where the additional degeneracy comes from the {\em
which layer} degree-of-freedom.  Although both groups used
non-linear sigma model descriptions they reached different
conclusions on the properties of skyrmion excitations in $SU(N)$
quantum Hall ferromagnets, and in particular on the minimum charge a
skyrmion can carry. By explicitly constructing the {\em microscopic}
wave functions for collective modes and skyrmions, we demonstrate
that the lowest-energy skyrmion excitations of the system carry
charge $\pm 1$.  This analysis also allows us to enumerate the
internal degrees of freedom associated with an individual skrymion
and predict the number of gapless collective modes present in
$SU(4)$ skyrmion lattice states.

Our paper is organized as follows.  In Section II we identify exact
eigenstates of $SU(N)$ quantum Hall ferromagnets at integer filling
factors which have broken symmetry.  We argue that these states are
the ground states for any physically sensible repulsive interaction
between the particles, and obtain exact results for their elementary
particle-hole excitations. We find that in the absence of symmetry
breaking perturbations there are $M(N-M)$ gapless collective modes
when $M$ of the $N$ members of a Landau-level multiplet are
occupied.  In Section III we discuss the properties of skyrmions in
graphene.  We find that for Coulombic electron-electron
interactions, skyrmions are much more robust in systems like
graphene with a Dirac band structure, than for the more familiar
parabolic band systems.  We predict that dense-skyrmion systems will
be ubiquitous in the quantum Hall regime of graphene, and that at
even integer fillings skyrmion lattice states will have four
branches of gapless collective modes, two with $k^{1}$ dispersion,
one with $k^{3/2}$ dispersion and one with $k^{2}$ dispersion.
Finally in Section IV we briefly present some elementary
considerations on broken symmetry states in graphene at fractional
filling factors. We conclude in Section V with a brief summary of
our results and some final comments.

\section{Symmetry Breaking Ground States and Collective Mode Spectra}

For simplicity we ignore disorder and mixing between different
Landau levels. To keep the discussion general, we assume electrons
have $N$ internal states and that the electron-electron interaction
is independent of these internal states, so that the Hamiltonian is
$SU(N)$ symmetric. Thus the total degeneracy of each Landau level
(LL), including the internal degeneracy, is $N\times N_\Phi$, where
$N_\Phi$ is the number of flux quanta enclosed in the system. We
start by considering the case where the filling factor of the
valence LL (with LL index $n$) is an integer:
\begin{equation}
\nu_n = N_e/N_\Phi=M \le N,
\end{equation}
where $N_e$ is the number of electrons occupying the valence LL. We
note that in the case of graphene $N=4$ and the four-fold degenerate
$n=0$ LL is half-filled (corresponding to $\nu_0 = 2$) at zero
doping; thus the Hall conductance is
\begin{equation}
\sigma_{xy}=\nu e^2/h
\end{equation}
with
\begin{equation}
\nu=4n-2+\nu_n.
\end{equation}
In the absence of interactions the quantum Hall effect occurs only
when each of the four-fold degenerate Landau levels is completely
full ($\nu_n = N = 4$) or completely empty ($\nu_n = 0$) and the
quantized Hall conductance is
\begin{equation}
\sigma_{xy}=(4 n + 2)e^2/h,
\end{equation}
where $n$ is the index of the highest {\em filled} Landau level.
Quantum Hall Ferromagnetism will\cite{nomura} lead to quantum Hall pleateaus
characterized by intermediate integers and by fractions.
In the following we neglect interaction induced mixing between
orbitals with different $n$.  For different values of $n$ the
properties of the $SU(N)$ quantum Hall ferromagnet are different.
In the following the dependence on $n$ is sometimes left implict
to avoid notational clutter.

Because of the $SU(N)$ symmetry of the system, the following single
Slater determinant state (in which only electronic states in the
valence LL are explicitly described) is an {\em exact} eigen state
of the Hamiltonian:
\begin{equation}
|\Psi_0\rangle =\prod_{1\le\sigma\le
M}\prod_{k}c_{k,\sigma}^\dagger|0\rangle. \label{ground}
\end{equation}
Here $c^\dagger$ is the electron creation operator, $|0\rangle$ is
the vacuum state, $\sigma$ is the index of the internal state that
runs from 1 to $N$, $k$ is an intra LL orbital index; for example in
the Landau gauge it is the wave vector along the plane wave
direction, while in the symmetric gauge it is the angular momentum
quantum number.

For a broad class of repulsive interactions, we expect
$|\Psi_0\rangle$ to be the exact {\em ground state} of the system;
for the case of $\nu_n=M=1$ this can be proved rigorously for
short-range repulsion (see below).  In this paper we use this
assumption as the starting point of our discussion.  Obviously, the
ground state $|\Psi_0\rangle$ breaks the $SU(N)$ symmetry
spontaneously since an $SU(N)$ rotation transforms the $M$
spontaneously chosen occupied single electron orbitals to another
different set. It therefore represents the ground state of an
$SU(N)$ ferromagnet, and is expected to support ferromagnetic spin
waves. In the following we construct the {\em exact} single spin
wave states and determine their spectrum in a manner similar to that
of Kallin and Halperin.\cite{kallin} Consider the following Landau
gauge states, with $\hat{y}$ in the direction of the plane waves:
\begin{eqnarray}
&&|{\bf
k}\rangle_{\sigma_1\sigma_2}=|k_x,k_y\rangle_{\sigma_1\sigma_2}\nonumber\\
&=&{1\over
\sqrt{N_\Phi}}\sum_{k'}e^{ik_xk'\ell^2}c^\dagger_{k'+k_y,\sigma_2}c_{k',\sigma_1}|\Psi_0\rangle.
\label{spinwave}
\end{eqnarray}
These spin-wave states are labeled by a two-dimensional wavevector
${\bf k}$ and two internal indices: $1\le \sigma_1 \le M$ and $M <
\sigma_2\le N$. $\ell=\sqrt{\hbar c/eB}$ is the magnetic length.
Physically they can be understood as single particle-hole states
formed from the $\sigma_2$ and $\sigma_1$ internal states, {\em
i.e.} as magnetic excitons. Since there are $M$ choices for
$\sigma_1$ and $N-M$ choices for $\sigma_2$, the total number of
these spin wave modes is $M(N-M)$. It follows from translational
invariance that ${\bf k}$ is a good quantum number, from $SU(N)$
invariance that excitons with distinct $(\sigma_2,\sigma_1)$ labels
are uncoupled, and that $|{\bf k}\rangle_{\sigma_1,\sigma_2}$
is therefore an exact eigenstate of the Hamiltonian. The
exact $SU(N)$ quantum Hall ferromagnet magnetoexciton dispersion relation is
\begin{eqnarray}
&&E(k)=\langle{\bf k}|\hat{V}|{\bf k}\rangle-\langle \Psi_0|\hat{V}|\Psi_0\rangle\nonumber\\
&=&{1\over
2\pi}\int_0^{\infty}qV(q)[F_n(q)]^2e^{-q^2\ell^2/2}[1-J_0(qk\ell^2)]dq,
\label{spectrum}
\end{eqnarray}
where $V(q)$ is the Fourier transform of electron-electron
interaction, $J_0$ is the Bessel function, and
\begin{equation}
F_n(q)={1\over 2}[L_{|n|}(q^2\ell^2/2)+L_{|n|-1}(q^2\ell^2/2)]
\label{eq:formfactor}
\end{equation}
is the Landau level structure factor appropriate for Dirac
fermions.\cite{nomura,goerbig} (In Eq. (\ref{eq:formfactor})
$L_n(x)$ is a Laguere polynomial.  The dependence of graphene
quantum Hall ferromagnet properties on $n$ enters only through
$F_n$.) The second term in Eq.~\ref{spectrum} is due to
particle-hole attraction\cite{kallin} and vanishes for $k \to
\infty$ where $E(k)$ approaches the energy of a widely separated
electron-hole pair.  For the case of a $1/r$ Coulomb interaction and
$n=0$ ($F_0(x) \equiv 1$), we have
\begin{eqnarray}
E(k)= {e^2\over \epsilon\ell}\sqrt{\pi\over
2}\left[1-e^{-k^2\ell^2/4}I_0(k^2\ell^2/4)\right],
\end{eqnarray}
where $\epsilon$ is the effective dielectric constant and $I_0$ is
the modified Bessel function. In the long-wave length limit
$E(k)\propto k^2$ as expected for ferromagnetic spin waves.

It is worth pointing out here that among the various sources of
perturbations that break the $SU(4)$ symmetry in graphene, the
simplest but most important one is the Zeeman splitting of electron
spin:
\begin{equation}
H_z=-g\mu_BBS_{tot}^Z,
\end{equation}
where $g$ is the electron spin $g$ factor and $\mu_B$ is the Bohr
magneton. Due to the fact that the total spin along the magnetic
field direction $S_{tot}^Z$ commutes with the $SU(4)$ invariant
Hamiltonian, it only shifts the energies of individual eigenstates
without changing the states themselves. In the presence of $H_Z$
electrons will choose to occupy spin-up states in the ground state
Eq. (\ref{ground}) until all such states are filled; the internal
states as labeled by $\sigma$ are thus eigen states of $S^Z$. The
single spin-wave states take the same form as in Eq.
(\ref{spinwave}); the spectrum remains the same as Eq.
(\ref{spectrum}) if $\sigma_1$ and $\sigma_2$ are of the same spin
orientation, while it gets shifted to $E(k)+g\mu_BB$ if they are of
opposite spin orientations.

Returning to the $SU(N)$ symmetric case, we note that the nature of
the symmetry breaking as realized in the ground state
(\ref{ground}), as well as the resultant gapless spin wave
excitations, may be understood from the following group-theoretical
analysis. The state (\ref{ground}), while not invariant under a
general $SU(N)$ transformation, is invariant under a subgroup of
$SU(N)$, $SU(M)\times SU(N-M)$. Physically the $SU(M)\times SU(N-M)$
subgroup represents $SU(M)$ transformations among the $M$ occupied
levels, and $SU(N-M)$ transformations among the $N-M$ unoccupied
levels; these transformations do {\em not} change the state
(\ref{ground}). Thus the order parameter as represented by the
symmetry breaking state (\ref{ground}) forms a coset space of
$SU(N)/SU(M)\times SU(N-M)\times U(1)$ (where the last $U(1)$
represents an overall phase difference between the occupied and
unoccupied levels), or equivalently, $U(N)/U(M)\times U(N-M)$. Since
a $U(N)$ transformation is parameterized by $N^2$ parameters (its
number of generators), we find the dimensionality (or the number of
independent fields) of the coset space to be
$N^2-M^2-(N-M)^2=2M(N-M)$.  Because we are dealing with a
ferromagnetic state (whose effective action contains a dynamic term
with a single time derivative, see below), half of these fields are
the conjugate momenta of the other half; we thus expect $M(N-M)$
independent spin-wave modes, in agreement with the microscopic
construction (\ref{spinwave}). Very similar analyses were performed
in Refs. \onlinecite{arovas,ezawa03} that lead to the same
conclusion.

The analysis above suggests the following matrix parametrization of
the ferromagnetic order parameter, appropriate for the symmetry
breaking corresponding to $U(N)/U(M)\times U(N-M)$\cite{arovas}:
\begin{equation}
Q({\bf r}, t)=U^\dagger({\bf r}, t)\hat{S}U({\bf r}, t),
\end{equation}
where $U({\bf r}, t)$ is the (space-time dependent) $N\times N$
$U(N)$ transformation matrix, $\hat{S}_{ij}\propto \delta_{ij}$ is a
diagonal matrix with $\hat{S}_{ii}=1$ for $0 < i \le M$ and
$\hat{S}_{ii}=-1$ for $M < i \le N$. The $N\times N$ matrix field
$Q({\bf r}, t)$ is the order parameter. Ezawa and
co-workers\cite{ezawa03} use a different, but presumably equivalent,
parametrization of the order parameter.

In terms of the matrix field $Q({\bf r}, t)$, the long-wave length,
low-energy effective action of the system takes the
form\cite{arovas}
\begin{eqnarray}
S[Q({\bf r}, t)]&=&\int{dtd{\bf r}}\left\{\alpha {\rm tr}
A(Q)\cdot\partial_tQ \right . \nonumber\\
&+& {1\over 4}\rho_s{\rm tr} (\nabla Q)\cdot(\nabla Q) + \left .
\cdots \right\}, \label{action}
\end{eqnarray}
where the first term is the Berry phase term\cite{arovas} that
encodes the commutation relations between different components of
the order parameter field, and the second term describes the energy
cost when the order parameter is non-uniform. The choice of the
prefactor $1/4$ (instead of $1/2$) for the second term is for later
convenience as it compensates for the fact that the trace of the
square of Pauli matrices is two. Terms involving higher orders of
either time or spatial derivatives, as well as possible symmetry
breaking perturbations are represented by $\cdots$. Thanks to the
knowledge of the exact ground state, the order parameter stiffness
$\rho_s$ may be determined exactly in a manner similar to that of
Ref. \onlinecite{moon}: We first construct a state by performing a
slow, ${\bf r}$-dependent $SU(N)$ rotation on $|\Psi_0\rangle$, then
project it to the appropriate LL, calculate its energy, and perform
a gradient expansion of the energy functional. The result is
\begin{equation}
\rho_s={1\over
32\pi^2}\int_0^{\infty}q^3V(q)[F_n(q)]^2e^{-q^2\ell^2/2}dq,
\label{eq:rhos}
\end{equation}
and for the case of $n=0$ and $1/r$ Coulomb interaction,
$\rho_s=e^2/(16\sqrt{2\pi}\epsilon\ell)$.\cite{arovas} $\rho_s$ is
independent of both $N$ and $M$; Eq.~(\ref{eq:rhos}) is identical to
the  $N=2$ and $\nu_n=M=1$ $\rho_s$ expression derived in earlier
work\cite{sondhi,moon}. This finding is not surprising since any
infinitesimal rotation in the $U(N)/U(M)\times U(N-M)$ subgroup of
$SU(N)$ can be decomposed into combinations of $SU(2)$ rotations
between occupied and unoccupied levels. It is easy to verify that
the action (\ref{action}) reproduces the spin wave spectrum
(\ref{spectrum}) in the long wave-length limit.

\section{Single Skyrmion States and Collective Modes of Skyrmion Lattices}

The matrix field $Q$ supports topologically non-trivial spatial
configurations, which can be parameterized by an integer-valued
topological quantum number called the Pontryagon index:
\begin{equation}
q={i\over 16\pi}\int{d^2{\bf r}}\epsilon^{\mu\nu}{\rm
tr}[Q\partial_\mu Q\partial_\nu Q],
\end{equation}
where $\epsilon^{\mu\nu}$ is the antisymmetric tensor. Field
configurations with nonzero $q$ carry topological charge and are
called skyrmions.  As in the $N=2$,\, $\nu_n=M=1$
case,\cite{sondhi,moon} quantum Hall ferromagnet skyrmions also
carry an {\em electric} charge that is equal to its topological
charge.  It follows from the above considerations, as concluded in
earlier work,\cite{arovas} that skyrmions with charge $\pm 1$ exist
within the lowest Landau level.  This remarkable property implies
that when skyrmions are the lowest energy charged excitations, they
will be present\cite{sondhi,fertig,skyrtheory,mfb} in the ground
state of the system when $\nu$ is close to, but not equal to, an
integer.  Skyrmions thus appear as emergent low-energy degrees of
freedom and influence all observable properties.

\begin{table}[ptb]%
\begin{tabular}
[c]{|c|c|c|c|c|}\hline
$\vert$ LL Index $\vert$& \; \; $\Delta_{QP}^{(D)}$ \;& $\; \Delta_{SK}^{(D)}$ \; & \; $\Delta_{QP}^{(P)}$ \; & \; $\Delta_{SK}^{(P)}$ \; \\\hline
0  & 1 & 1/2 & 1 & 1/2 \\\hline
1  & 11/16 & 7/32 & 3/4 & 7/8  \\
& (0.6875) & (0.2188) & (0.75) & (0.875) \\\hline
2  & 145/256 & 169/512 & 41/64 & 145/128 \\
& (0.5664) & (0. 3301) & (0.6406) & (1.1328) \\\hline
3  &\; 515/1024 \;&\; 839/2048 \;&\; 147/256 \;&\; 687/512 \;\\
& (0.5029) & (0.4097) & (0.5742) & (1.3418) \\\hline
4  & 0.4608 & 0.4754 & 0.5279 & 1.5522 \\\hline
5  & 0.4298 & 0.5328 & 0.4927 & 1.6834 \\\hline
\end{tabular}
\caption{Hartree-Fock Quasiparticle ($\Delta_{QP}$) and Skyrmion/Antiskrymion ($\Delta_{SK}$) particle-hole
excitation gaps for the Dirac $(D)$ bands of graphene and for parabolic bands $(P)$.  The two cases are
distinquished by different dependences of form factor $F_n$ on Landau level (LL) index $n$.  These results are
for Coulomb interactions in an ideal two-dimensional electron system without finite thickness corrections
and energies are in units of $e^2/\epsilon \ell \; \sqrt{\pi/2}$.  In graphene the effective value of $\epsilon$
depends on the dielectric screening environment provided by the substrate but is typically less than $2$, whereas
in GaAs and other common heterojunction systems $\epsilon \sim 10$.}%
\label{gaps}%
\end{table}

To determine whether or not skyrmions are the lowest energy charged
excitation for a paraticular $n$ we need to compare the energy of a
skyrmion/antiskyrmion pair,
\begin{equation}
\Delta_{SK} = 8 \pi \rho_s,
\end{equation}
with the energy of an ordinary Hartree-Fock theory particle-hole pair,
\begin{equation}
\Delta_{PH} = E(k \to \infty).
\end{equation}
Table ~\ref{gaps} compares results for graphene Dirac-band and the
ordinary parabolic band cases.  In the Dirac-band cases both
positive and negative values of $n$ occur whereas the Landau level
indices of parabolic systems are non-negative integers. In both
cases, the increase in cyclotron orbit radius with $|n|$ is
reflected in the quantum form factor $F_n$; for the parabolic case
the form factor $F_n = L_n^2(q^2\ell^2/2)$.  Since the cyclotron
orbit radius $R_c \sim \ell \sqrt{n} $ it is clear simply on
dimensional grounds that for Coulomb interactions and large $n$,
$\Delta_{PH} \sim e^2/R_c \sim e^2/\ell \times 1/\sqrt{n}$ whereas
$\Delta_{SK} \sim e^2 R_c / \ell^2 \sim e^2/\ell \times \sqrt{n}$.
This difference in the large $n$ behavior is already apparent in
both cases in Table ~\ref{gaps}.  For sufficiently large $n$ then,
$\Delta_{SK}$ will exceed $\Delta_{QP}$, the lowest energy charged
excitations will be ordinary Hartree-Fock quasiparticles, and the
ground state near integer filling factors will not have low-energy
Skyrmion degrees of freedom.  The quantitative calculations
summarized in Table ~\ref{gaps} show that ordinary quasiparticles
are already energetically preferred for $n=1$ in the parabolic band
case, a result obtained first by Wu and Sondhi.\cite{sondhiwu}
Interestingly the crossover to ordinary quasiparticles does not
occur until $n=4$ in the Dirac band case; thus skyrmion physics will
occur within $n=0$, $n=1$, $n=2$ and $n=3$ Landau levels in
graphene.

We note in passing that we have so far compared only
quasiparticle/quasihole and skyrmion/antiskyrmion pair excitation
energies. In order to conclude that skyrmions and antiskyrmions are
present in the ground state on both sides of integer filling
factors, we need to demonstrate that the skyrmion energy and the
antiskyrmion energy are separately smaller than the quasiparticle
and quasihole  energies respectively, whenever the pair excitation
energies are smaller. In the case of $N=2$ and $\nu_n=M=1$ this
property follows\cite{elandholeenergy} from particle-hole symmetry.
We demonstrate below that the pair excitation energy criterion also
applies for graphene, although the justification is subtly different
for $\nu_n=1$ and $\nu_n=3$ cases compared to the $\nu_n=2$ case.

In earlier work Ezawa\cite{ezawa03} and co-workers concluded that
for $\nu_n=M > 1$, lowest Landau level (LLL) skyrmions must have a
charge that is a multiple of $M$. To address this discrepancy, we
explicitly construct LLL skyrmion wave functions and demonstrate
that it is indeed possible to have charge $\pm 1$.  Our microscopic
single Slater determinant skyrmion wave function is constructed in a
manner similar to that of Ref. \onlinecite{fertig}.  For
definiteness we discuss a hole-like skyrmion with $q=-1$:
\begin{eqnarray}
|\Psi_{sky}\rangle&=&\prod_{m=0}^{N_\Phi
-2} \, \left[u_m c^\dagger_{M,m+1}+ v_{m} c^\dagger_{M+1,m} \right ] \nonumber\\
&&\prod_{m=0}^{N_\Phi -1}\left[ \prod_{1 \le \sigma < M} \,
c^\dagger_{\sigma,m}\right]|0\rangle, \label{sky}
\end{eqnarray}
where the internal state labels match those of the symmetry broken
ground state. Hartree-Fock skyrmion states are obtained by
minimizing the expectation value of the Hamiltonian with respect to
$u_m$ and $v_m$. Skyrmion states with larger topological charges may
be constructed in a similar manner. These single Slater determinant
(or Hartree-Fock) skyrmion states correspond to classical skyrmions
of the field theory (\ref{action}), although the Coulomb
self-interaction energy of skyrmions must be included in the field
theory\cite{sondhi} in order to describe small skyrmions.

The Hartree-Fock (or semi-classical) skyrmion state (\ref{sky}) has
a rich internal structure, which may be analyzed in a way similar to
the analysis performed earlier on the ground state (\ref{ground}).
The state (\ref{sky}) is invariant only under unitary
transformations among the first $M-1$ labels or the final $N-M-1$
labels.  It follows that the family of $SU(N)$ transformations that
actually transforms the skyrmion states form a coset space of
dimension $N^2 - (N-M-1)^2 - (M-1)^2 = 2M(N-M) + 2 (N-1)$. The set
of transformations of dimension $2M(N-M)$ corresponds to the order
parameter fields of the ground state itself, and the additional $2
(N-1)$-dimensional space to $(N-1)$ skyrmion internal complex
degrees-of-freedom.  These must be specified in addition to location
and size (determined by minimizing the energy with respect to $u_m$
and $v_m$) in order to completely characterize a classical skyrmion.
In other words, the skyrmion has $N-1$ internal (complex) degrees of
freedom. A more intuitive way to understand this point is the
following. Let us assume the ground state order parameter
configuration is fixed (either spontaneously or by symmetry breaking
pertubations). We now introduce a single skyrmion at the origin, and
minimize its energy. This fixes the size of the skyrmion, and the
{\em magnitudes} of $u_m$ and $v_m$. On the other hand $u_m$ is a
[$CP(M-1)$] vector that lives in an $SU(M)$ space spanned by the $M$
occupied levels of the ground state, so there are $M-1$ remaining
internal (complex) degrees of freedom associated with it. Similarly
there are $N-M-1$ remaining internal degrees of freedom associated
with $v_m$. Adding the relative overall phase between $u_m$ and
$v_m$, we find the total internal degrees of freedom is $N-1$. As a
matter of fact $u_m$ and $v_m$ may be combined and viewed as an
$CP(N-1)$ vector in the original $SU(N)$ space; thus the $N-1$
internal (complex) degrees of freedom of a skyrmion may be viewed as
those of a $CP(N-1)$ superspin.\cite{liu}

Once we move away from $\nu_n=M$, we have a finite density of
skyrmions in the ground state, provided that they are indeed the
lowest energy charged excitations.  For graphene we are able to
judge the energetic ordering of charged excitations based on the
energetic ordering of particle-hole excitations.  To justify this
statement at $\nu_n=M=1$, we must appeal to Zeeman coupling which is
always present experimentally.  Zeeman coupling selects a fully
spin-polarized state and also selects fully spin-polarized
skyrmions. We can therefore simply ignore the spin-degree of freedom
and appeal to the same arguments\cite{elandholeenergy} used for
$\nu_n=M=1$ when $N=2$.  For $\nu_n=M=3$ we can then appeal to
particle-hole symmetry in the $N=4$ Landau level multiplet, which
suggests that the situation is the same as $\nu_n=M=1$. Finally for
$\nu_n=M=2$ we can generalize the argument of
Ref.~\onlinecite{elandholeenergy} by appealing directly to
particle-hole symmetry at this filling.

In their classical ground state skyrmions will form a lattice.
(Quantum corrections to the classical ground state become more
important as the skyrmion density increases.\cite{skyrnmr,nazarov})
In this case the internal degrees of freedom associated with
skyrmions on different sites will interact and the classical energy
will be minimized by an arrangement with long-range order. When
quantized, fluctuations in the internal degrees of freedom will
result in $N-1$ spin wave-like modes, which will be present in
addition to the single phonon mode associated with fluctuation in
skyrmion positions. (In a strong magnetic field transverse and
longitudinal position fluctuations are canonically conjugate
leading\cite{magnetophonon} to phonons with $k^{3/2}$ dispersion.)
In the presence of full $SU(N)$ symmetry these internal modes are
gapless, and some (or all) of them may remain gapless in the
presence of symmetry-breaking perturbations, under appropriate
conditions (see below for examples). For the $SU(2)$ case we find
$N-1=1$ internal mode, which is known previously;\cite{skyrnmr} here
we provide a more general understanding of this result.

For graphene $N=4$, and the case $\nu_n=M=2$ is particularly
interesting. Weak Zeeman coupling will select a {\em unique} fully
spin-polarized, valley-singlet ground state.  All collective modes
are therefore gapped for $\nu_n$ exactly equal to $2$, {\em i.e.} in
the absence of skyrmions. In this case the $SU(2)$ valley symmetry
is {\em preserved} by the ground state. We nevertheless predict that
the skyrmion lattice state near $\nu_n=2$ will have phonons and $3$
additional {\em gapless} modes. Of these three internal modes, we
predict that one has {\em quadratic} dispersion, while the other two
will have linear dispersions. The quadratic mode may be understood
in the following manner. The skyrmion lattice state has a finite
pseudospin (valley) magnetization, and is thus a spontaneous valley
ferromagnet; the quadratic mode is simply the ferromagnetic spin
wave in the valley channel. The two linear modes are the Goldstone
modes corresponding to the two additional spontaneously broken
$U(1)$ symmetries; these symmetries are broken by the fixed relative
phases between $u_m$ and $v_m$ with the same valley index. Adding
the phonon mode, we thus find one quadratic mode, one $k^{3/2}$
mode, and two linear modes, all gapless.

The situation is very different at $\nu_n=M=1$. In this case, the
ground state is still fully spin-polarized due to the Zeeman
coupling, but is also a {\em spontaneous} valley ferromagnet that
breaks the $SU(2)$ valley symmetry spontaneously. Thus we expect a
single gapless mode with quadratic dispersion in the absence of
skyrmions, the $SU(2)$ valley pseudospin waves. The lowest energy
skyrmion states are thus pseudospin textures in the valley degree of
freedom,\cite{mansour} and the skyrmion lattice states at low
temperature will be fully spin-polarized. As a result, we expect
only one gapless internal mode with quadratic dispersion which does
not involve spin flips, while the other two internal modes which do
involve spin flip will have a gap of $g\mu_BB$. Thus the only new
gapless mode of the skyrmion lattice state is the phonon mode in
this case.

Isolated individual {\em quantum} skyrmion states have well-defined
total orbital angular momentum and internal $SU(N)$ quantum numbers,
which can be analyzed for $\nu_n=M=1$ following the strategy of Ref.
\onlinecite{mfb}. These are the quantum counterparts of classical
$CP(N-1)$ skyrmions. In this case the single Slater determinant
state (\ref{ground}) is the exact ground state for $\delta$-function
interaction between electrons, since it has exactly zero energy.
Hole-like skyrmion states may be identified as all zero energy
states for the case $N_\Phi=N_e+1$. As emphasized in Ref.
\onlinecite{mfb} (see also Ref. \onlinecite{tsit}), all such states
may be written in the form
\begin{equation}
\Psi(z,\chi)=\left[\prod_{i<j}(z_i-z_j)\right]\Psi_B(z,\chi),
\label{qmsky}
\end{equation}
where the antisymmetric Jastrow factor $\prod_{i<j}(z_i-z_j)$
ensures zero energy, while $\Psi_B(z,\chi)$ is a {\em bosonic} wave
function that is {\em symmetric} under the exchange of spatial and
internal coordinates of two particles. We can classify the skyrmion
states based on the properties of this bosonic wave function.

Due to the fact $N_\Phi=N_e+1$, the bosons can only be in $m=0$ or
$m=1$ orbital states.  Letting these occupation numbers be $n_0$ and
$n_1$ respectively, the total angular momentum of a state
(\ref{qmsky}) measured from that of the ground state is $\Delta L =
n_1$, thus $n_0=N_e - \Delta L$. We now classify all the skyrmion
states for a fixed $\Delta L$ based on the $SU(N)$ representations
they form (in the familiar $SU(2)$ case these representations are
labeled by a single quantum number, the total spin\cite{mfb}).
Because bosons occupying the same orbital have totally symmetric
orbital wave functions, their internal wave function must also be
totally symmetric. It follows that the internal states of the $n_0$
bosons in orbital $m=0$ form a totally symmetric representation of
$SU(N)$, as do the $n_1$ bosons in orbital $m=1$. In terms of Young
tableau, they form two horizontal row representations, with $n_0$
and $n_1$ horizontal boxes respectively. More generally, each
irreducible representation of the $SU(N)$ group as represented by
the Young tableau\cite{frappat} can be labeled by a set of $N-1$
non-negative integers in descending order: $[l_1, l_2, \cdots,
l_{N-1}]$, where each integer represent the number of boxes in each
row. Thus the two representations formed by the bosons in $m=0$ and
$m=1$ orbitals form representations $[n_0, 0, \cdots]$ and $[n_1, 0,
\cdots]$ respectively. Now we take the direct product of these two
representations and decompose them into irreducible representations
of $SU(N)$:
\begin{eqnarray}
&[&n_0, 0, \cdots]\otimes [n_1, 0, \cdots] =
[n_0+n_1,0,0,\cdots]\nonumber\\
 &\oplus& [n_0+n_1-1,1,0,\cdots]\oplus\cdots\oplus[n_0,n_1,0,\cdots]\nonumber\\
 &=&[N_e,0,0,\cdots]\oplus[N_e-1,1,0,\cdots]\oplus\cdots\nonumber\\
 &\oplus&[N_e-\Delta L,\Delta L,0,\cdots].
\end{eqnarray}
In the above we have assumed that $n_0\ge n_1$, or $N_e\ge 2\Delta
L$. If the opposite is true the positions of $n_0$ and $n_1$ need to
be interchanged. The dimensionality of these representations may be
found in the literature\cite{frappat}. This procedure classifies the
hole-like skyrmion states at $\nu_n=1$ based on their angular
momentum quantum number and the irreducible $SU(N)$ representations
they form.

\section{SU(N) Ferromagnets at Fractional Filling Factors}

We now turn our discussion to possible fractional quantum Hall (FQH)
states (which have not yet been observed in graphene), many of which
are also $SU(N)$ ferromagnets. Many FQH states may be constructed by
starting with integer quantum Hall states\cite{jain} and using the
composite fermion (CF) flux-attachment {\em ansatz}. In this
construction, we start from an IQH state with filling factor
$\nu_{n,CF}=m$, and attach an even $2n'$ flux quanta to the CFs to
turn them into electrons; after the flux is spread out the electron
filling factor in the valence LL becomes
\begin{equation}
\nu_n={m\over 2n'm\pm 1}. \label{fqh}
\end{equation}
This is, of course, the familiar Jain's sequence.\cite{jain,peres}
The difference here is that in the presence of the internal
degeneracy and $SU(N)$ symmetry, the CF IQH state is an $SU(N)$
ferromagnet as long as $m$ is not a multiple of $N$. We thus expect
the $SU(N)$ symmetry properties to be reflected in the FQH states.
The number of collective modes and the number of branches of
skyrmion excitations should be the same as that of the IQH states at
$\nu_n=M$, if we identify $M={\rm mod}(m, N)$. On the other hand if
$m$ is a multiple of $N$, we obtain an $SU(N)$ singlet, and there is
no spontaneous symmetry breaking; the system will be fully gapped.
For cases with $m\le N$, we write down below Laughlin-Halperin type
of trial wave functions for the FQH states with the expected $SU(N)$
symmetry properties in first quantization:
\begin{equation}
\psi_{n',m}(z)=\left[\prod_{i<j}^{N_e}(z_i-z_j)^{2n'}\right]\left[A\prod_{\sigma=1}^m\prod_{k<l}^{N_\Phi}(z_{k\sigma}-z_{l\sigma})\right].
 \label{fqhwf}
\end{equation}
Here $A$ represents antisymmetrization of all coordinates, and we
have neglected the common exponential factors for LL wave functions.
The second factor is the first quantized wave function for
(\ref{ground}), while the first factor reflects flux attachment. The
wave functions for $m > N$ is more complicated, as in this case some
of the CFs occupy higher LLs, and LLL projection is
necessary\cite{jain}. We note that this type of construction was found to
be generally reliable in predicting the spin structure in the $SU(2)$
case without Zeeman splitting\cite{jain}.

The low-energy physics of the FQH $SU(N)$ ferromagnets are also
described by field theory (\ref{action}). In this case we do not
have exact knowledge about the parameters (like the stiffness
$\rho_s$) of the theory, but $\rho_s$ has been calculated
numerically for the members of the Laughlin sequence $\nu=1/3, 1/5$
for $1/r$ interaction based on the Laughlin trial wave function in
the $SU(2)$ case\cite{moon}; they are $9.23\times
10^{-4}(e^2/\epsilon\ell)$ and $2.34\times
10^{-4}(e^2/\epsilon\ell)$ respectively. Using the same arguments as
for the IQH case discussed above, we expect $\rho_s$ to take the
same two values at the same filling factors in graphene. These
values can be used to determine the energies of
skyrmion-antiskyrmion pairs, which may be compared with transport
measurements should FQH states be observed in graphene in future
experiments.  We caution however that at fractional filling factors,
we can no longer appeal to particle-hole symmetry properties so
that it is possible\cite{su2case} in general for Laughlin-like fractionally charged
quasiparticles to be present in the ground state on one side of
an incompressible filling factor and fractionally charged skyrmions
on the other.

\section{Concluding Remarks}

In closing we comment on the possible effect of single-particle
valley splitting on our $SU(4)$-based analysis of graphene. We first
note that the valley degeneracy of graphene is rather robust; for
example unlike in Silicon, simple strain cannot lift the degeneracy.
Electron-electron interactions are likely to provide the most
important source of Hamiltonian matrix elements that break the
valley portion of the $SU(4)$ symmetry.  Our considerations should
nevertheless largely apply for weak valley-symmetry breaking, as
long as the characteristic energy scale of these terms is much
weaker than the Coulomb interaction scale. We mention that among the
possible extrinsic sources of 4-fold degeneracy lifting in the
single-particle Hamiltonian are edge effects and inter-valley
scattering. For large extrinsic splittings, the $SU(4)$ symmetry
will be reduced to $SU(2)\times U(1)\times U(1)$, where the
remaining $SU(2)$ symmetry corresponds to spin, which is further
reduced by Zeeman splitting as discussed earlier. At the lowest
temperatures and energy scales the physics may be be similar to that
of a bilayer system\cite{jungwirth}, which may support other types
of broken symmetry states\cite{canted,rajaraman}. The question of
whether our predicted interaction-driven spontaneous breaking of
valley degeneracy is playing the key role in the observed valley
splitting at high field\cite{zhang2} can be decided by careful
measurements of zero-field valley splitting.  If the experimental
zero-field valley splitting is negligibly small, then it seems
certain that $SU(4)$ quantum Hall ferromagnetism, associated with
the spontaneous breaking of valley and spin degeneracy, is already
playing a role in the high-field quantum Hall experiments in
graphene.  (It seems clear that many-body physics does not have a
large influence on quasiparticle valley-splitting in the absence of
a field.) Direct observations of skyrmions and associated collective
excitations in graphene then take on particular experimental
relevance. We further note that both Zeeman and valley splittings
are single electron effects; electron-electron interactions also
have weak symmetry breaking effects due physics at lattice scale, as
discussed recently.\cite{goerbig}

In summary, we have identified exact broken symmetry eigenstates of
$SU(N)$-invariant Hamiltonians in the quantum Hall regime of
graphene. We argue that these states are ground states for any
physically sensible repulsive interaction between the particles, and
for Coulomb interactions between electrons in particular. Given
$SU(N)$ invariance we were able to obtain exact results for the
elementary collective excitation spectrum. We found that in the
absence of symmetry breaking there are $M(N-M)$ gapless collective
modes when $M$ of the $N$ members of a Landau-level multiplet are
occupied.  In addition, we have shown that for Coulombic
electron-electron interactions, skyrmions are much more robust in
systems like graphene with a Dirac band structure, than for the more
familiar parabolic band systems. We predict that dense-skyrmion
systems will be ubiquitous in the quantum Hall regime of graphene
and that skyrmion lattice states near even integer filling factors
will have four branches of gapless collective modes, two with
$k^{1}$ dispersion, one with $k^{3/2}$ dispersion and one with
$k^{2}$ dispersion. The identification of probes that can study
skyrmion physics in graphene layers is therefore an attractive
challenge for experiment.

\acknowledgements The authors acknowledge useful discussions with
Shou-Cheng Zhang, and the hospitality of Kavli Institute for
Theoretical Physics where this work was performed. This work was
supported by National Science Foundation grant DMR-0225698 and a
Florida State University Research Foundation Cornerstone grant (KY),
by US-ONR and LPS-NSA (SDS), and by the Welch Foundation (AHM).

{\em Note added} --  While this paper was being written, a
preprint\cite{alicea} authored by Alicea and Fisher
appeared which has some overlap with the present paper.
Among other contributions, Alicea and Fisher\cite{alicea} studied
collective modes in the long-wave length limit, and single skyrmion
states at certain specific integer fillings.  Where overlap occurs,
our results are consistent with those of Alicea and Fisher.
We thank Jason Alicea for informative conversations about this work.


\begin{thebibliography}{1}

\bibitem{graphene_exp} K. S. Novoselov {\em et al.}, Science {\bf 306}, 666 (2004);
Y. Zhang {\em et al.}, Phys. Rev. Lett. {\b 94}, 176803 (2005); C.
Berger {\em et al.}, J. Phys. Chem B {\bf 108}, 19912 (2004).

\bibitem{novoselov} K. S. Novoselov, A. K. Geim, S. V. Morozov, D. Jiang, M. I. Katsnelson, I. V. Grigorieva,
S. V. Dubonos and A. A. Firsov, Nature {\bf 438}, 197 (2005).

\bibitem{zhang} Yuanbo Zhang, Yan-Wen Tan, Horst L. Stormer and Philip Kim, Nature {\bf 438}, 201 (2005).

\bibitem{zhang2} Y. Zhang, Z. Jiang, J. P. Small, M. S. Purewal, Y.-W. Tan, M. Fazlollahi, J. D. Chudow,
J. A. Jaszczak, H. L. Stormer, P. Kim, Phys. Rev. Lett. {\bf 96},
136806 (2006).

\bibitem{nomura} K. Nomura and A. H. MacDonald, cond-mat/0604113.

\bibitem{sds}  S. Das Sarma and A. Pinczuk, eds. {\em Perspectives in
Quantum Hall Effects}, John Wiley and Sons, New York, 1997.

\bibitem{gm} For reviews, see articles by S. M. Girvin and A. H. MacDonald, and by J. P.
Eisenstein in Ref. \onlinecite{sds}.

\bibitem{sondhi} S. L. Sondhi, A. Karlhede, S. A. Kivelson, and E. H. Rezayi, Phys. Rev. B
{\bf 47}, 16419 (1993).

\bibitem{fertig} H. A. Fertig, L. Brey, R. Cote, and A. H.
MacDonald, Phys. Rev. B {\bf 50}, 11018 (1994).

\bibitem{skyrtheory}
L. Brey, H.A. Fertig, R. Cote, and A.H. MacDonald, Phys. Rev. Lett.
{\bf 75}, 2562 (1995); K. Yang and S. L. Sondhi, Phys. Rev. B {\bf
54}, R2331 (1996); E. H. Rezayi and S.L. Sondhi, Int. J. Mod. Phys.
{\bf 13}, 2257 (1999); J. Sinova, S.M. Girvin, T. Jungwirth, and K.
Moon, Phys. Rev. B {\bf 61}, 2749 (2000); A.G. Green, Phys. Rev. B
{\bf 61}, R16299 (2000); S. Sankararaman and R. Shankar, Phys. Rev.
B {\bf 67}, 245102 (2003); A.V. Ferrer, R.L. Doretto, and A.O.
Caldeira, Phys. Rev. B {\bf 70}, 045319 (2004); O. Bar, M. Imboden,
and U.J. Wiese, Nuc. Phys. B {\bf 686}, 347 (2004).

\bibitem{skyrexpt} S. E. Barrett, G. Dabbagh, L. N. Pfeiffer, K. W. West, and R. Tycko
Phys. Rev. Lett. {\bf 74}, 5112 (1995); A. Schmeller, J.P.
Eisenstein, L.N. Pfeiffer, and K.W. West, Phys. Rev. Lett. {\bf 75},
4290 (1995); E.H. Aifer, B.B. Goldberg, D.A. Broido, Phys. Rev.
Lett. {\bf 76}, 680 (1996); V. Bayot, E. Grivei, S. Melinte, M.B.
Santos, and M. Shayegan, Phys. Rev. Lett. {\bf 76}, 4584 (1996); V.
Bayot, E. Grivei, J.-M. Beuken, S. Melinte, and M. Shayegan, Phys.
Rev. Lett. {\bf 79}, 1718 (1997); D.R. Leadley, R.J. Nicholas, D.K.
Maude, A.N. Utjuzh, J.C. Portal, J.J. Harris, and C.T. Foxon, Phys.
Rev. Lett. {\bf 79}, 4246 (1997); J.L. Osborne, A.J. Shields, M.Y.
Simmons, N.R. Cooper, D.A. Ritchie, and M. Pepper, Phys. Rev. B {\bf
58}, R4227 (1998); S. Melinte, E. Grivei, V. Bayot, and M. Shayegan,
Phys. Rev. Lett. {\bf 82}, 2764 (1999); J.H. Smet, R.A. Deutschmann,
F. Ertl, W. der Wegschei, G. Abstreiter and K. von Klitzing, Phys.
Rev. Lett. {\bf 92}, 086802 (2004); P.G. Gervais, H.L. Stormer, D.C.
Tsui, P.L. Kuhns, W.G. Moulton, A.P. Reyes, L.N. Pfeiffer, K.W.
Baldwin, and K.W. West, Phys. Rev. Lett. {\bf 94}, 196803 (2005).

\bibitem{skyrnmr} R. Cote, A.H. MacDonald, L. Brey, H.A. Fertig,
S.M. Girvin, and H.T.C. Stoof, Phys. Rev. Lett. {\bf 78}, 4825 (1997).

\bibitem{sondhiwu} X.-G. Wu and S.L. Sondhi, Phys. Rev. B {\bf 51}, 14725 (1995).

\bibitem{goerbig} M.O. Goerbig, R. Moessner, and B. Doucot, cond-mat/0604554.

\bibitem{arovas} D. P. Arovas, A. Karlhede and D. Lilliehook, Phys. Rev. B {\bf 59},
13147 (1999).

\bibitem{ezawa03} K. Hasebe and Z. F. Ezawa, Phys. Rev. B
{\bf 66}, 155318 (2002); Z. F. Ezawa, G. Tsitsishvili, and K.
Hasebe, Phys. Rev. B {\bf 67}, 125314 (2003).

\bibitem{ezawa05} Z. F. Ezawa and G. Tsitsishvili, Phys.
Rev. D {\bf 72}, 085002 (2005).

\bibitem{kallin} C. Kallin and B. I. Halperin, Phys. Rev. B {\bf 30}, 5655
(1984).

\bibitem{moon} K. Moon, H. Mori, K. Yang, S. M. Girvin, A. H. MacDonald,
L. Zheng, D. Yoshioka and S.-C. Zhang, Phys. Rev. B {\bf 51}, 5138
(1995); K. Yang, K. Moon, L. Belkhir, H. Mori, S. M. Girvin, A. H.
MacDonald, L. Zheng and D. Yoshioka, Phys. Rev. B {\bf 54}, 11644
(1996); K. Yang, K. Moon, L. Zheng, A. H. MacDonald, S. M. Girvin,
D. Yoshioka, and Shou-Cheng Zhang, Phys. Rev. Lett. {\bf 72}, 732
(1994).

\bibitem{mfb} A. H. MacDonald, H. A. Fertig, and Luis Brey, Phys. Rev. Lett. {\bf 76}, 2153
(1996).

\bibitem{elandholeenergy} H.A. Fertig, L. Brey, R. Cote, A.H. MacDonald, A. Karlhede, and
S.L. Sondhi, Phys. Rev. B {\bf 55}, 10671 (1997).

\bibitem{nazarov}  Y.V. Nazarov and A.V. Khaetskii,
Phys. Rev. Lett. {\bf 80}, 576 (1998).

\bibitem{magnetophonon} Lynn Bonsall and A.A. Maradudin, Phys. Rev. B {\bf 15}, 1959 (1977).

\bibitem{liu} For the special case of $M=1$, the same conclusion has been obtained
by Liu and co-workers using different arguments; see X. Liu, Y.-S.
Duan, and P.-M. Zhang, Commun. Theor. Phys. (Beijing, China) {\bf
44}, 371 (2005).

\bibitem{mansour} Observations of such "valley skyrmions" has been
reported recently in silicon systems: Y.P. Shkolnikov, S. Misra,
N.C. Bishop, E.P. De Poortere, and M. Shayegan, Phys. Rev. Lett.
{\bf 95}, 066809 (2005).

\bibitem{tsit} G. Tsitsishvili and Z. F. Ezawa, Phys. Rev. B
{\bf 72}, 115306 (2005).

\bibitem{frappat} See, {\em e.g.}, L. Frappat, A. Sciarrino, and P.
Sorba, {\em Dictionary on Lie Algebras and Superalgebras}, Academic
Press, New York, 2000.

\bibitem{jain} For a review, see J. K. Jain, article in Ref. \onlinecite{sds}.

\bibitem{peres} Peres {\em et al.} [N. M. R. Peres, F. Guinea, and A. H. Castro Neto, Phys. Rev. B {\bf 73}, 125411 (2006)]
have recently constructed a different set of FQH states using flux attachment.

\bibitem{su2case} J.J. Palacios and A.H. MacDonald, Phys. Rev. B {\bf 58}, R10171 (1998);
A. Wojs and J.J. Quinn, Solid State Commun. {\bf 122}, 407 (2002).

\bibitem{jungwirth} For an analysis of the microscopic physics of $SU(2)$ symmetry breaking
in the $N=2$ case see T. Jungwirth and A.H. MacDonald, Phys. Rev. B {\bf 63}, 035305 (2001).

\bibitem{canted} S. Das Sarma, S. Sachdev, and L. Zheng, Phys. Rev. B {\bf 58}, 4672 (1998).

\bibitem{rajaraman} A. H. MacDonald, R. Rajaraman, and T. Jungwirth
Phys. Rev. B {\bf 60}, 8817 (1999).

\bibitem{alicea} J. Alicea and M. P. A. Fisher, cond-mat/0604601.

\end{thebibliography}
\end{document}